\documentclass[prl,floatfix,twocolumn,showpacs]{revtex4-1}

\usepackage{graphicx}
\usepackage{subfigure}
\usepackage{amsmath}
\usepackage{amssymb}
\usepackage{latexsym}
\usepackage{multirow}
\usepackage[colorlinks=true]{hyperref}

\begin{document}

\title{Instabilities of monolayered epithelia: shape and structure of villi and crypts}

\author{E. Hannezo$^1$, J. Prost$^{1,2}$, J.-F. Joanny$^1$}
\affiliation{$^1$Physicochimie Curie (Institut Curie / CNRS-UMR168 /UPMC), Institut Curie, 
Centre de Recherche, 
26 rue d'Ulm 75248 Paris Cedex 05 France}
\affiliation{$^2$E.S.P.C.I, 10 rue Vauquelin, 75231 Paris Cedex 05, France}

\date{\today}

\begin{abstract}

We study theoretically the shapes of a dividing  epithelial monolayer of cells lying on top of  an elastic stroma. The negative tension created by cell division provokes a buckling instability at a finite wave vector leading to the formation of periodic arrays of villi and crypts. The instability is similar to the buckling of a metallic plate under compression. We use the results to rationalize the various structures of the intestinal lining observed \emph{in vivo}. Taking into account the coupling between cell division and local curvature,  we obtain different patterns of villi and crypts, which could explain the different morphologies of the small intestine and the colon.
\end{abstract}
\pacs{xx}

\maketitle

An essential  property of  living tissues is a permanent cell turnover 
due to  division and apoptosis,  which 
has important effects on their mechanical response. Since cells often grow and 
divide in a constrained environment, cell division and cell death induce internal stresses
in tissues that influence deeply 
their architectures and morphologies. Conversely, there is a mechanical feedback 
on cellular growth, differentiation, and organ development, which is nowadays an active 
area of research \cite{1,2,3,9,10,32}. 

The intestine is the body's fastest renewing organ, and it is therefore a particularly 
interesting example to consider. It exhibits a variety of folded multicellular 
structures \cite{4}, called villi, which play a crucial role in favoring  the 
exchange of nutrients. Although the development of the intestine has been 
studied for various animals, the formation of villi and their structure 
has not yet been described quantitatively \cite{29}. 
From a physicist point of view, these folded shapes in a periodic 
arrangement are strongly reminiscent 
of the patterns observed after the buckling of metallic plates.
Euler buckling is the instability leading to the lateral deflection of an elastic beam or a 
surface under load. A similar buckling instability is expected in growing 
constrained systems \cite{13}, and has been invoked to explain fingerprints \cite{6} or the shapes of algae \cite{7}.
Of particular interest is the recent report \cite{8} that intestinal structures 
can be reproduced in vitro with minimal physiological environment, reinforcing 
the idea that mechanical forces play a crucial role. Still, studies of cell renewal in the intestine take villi shapes as a given phenomenological curve \cite{30}. We argue that it is in fact a consequence of cell renewal itself.

We present here a theoretical model for the intestinal structure based 
on the buckling instability of monolayered epithelial cells. We show 
that a few minimal ingredients are sufficient to capture many key features 
of the intestinal architecture. 
This model reproduces the correct orders of magnitude, the various patterns of 
folded structures observed \emph{in vivo} and gives insight into the 
physiological distinction between small intestine and colon.

\begin{figure}[!h]
\includegraphics[width=7.8cm]{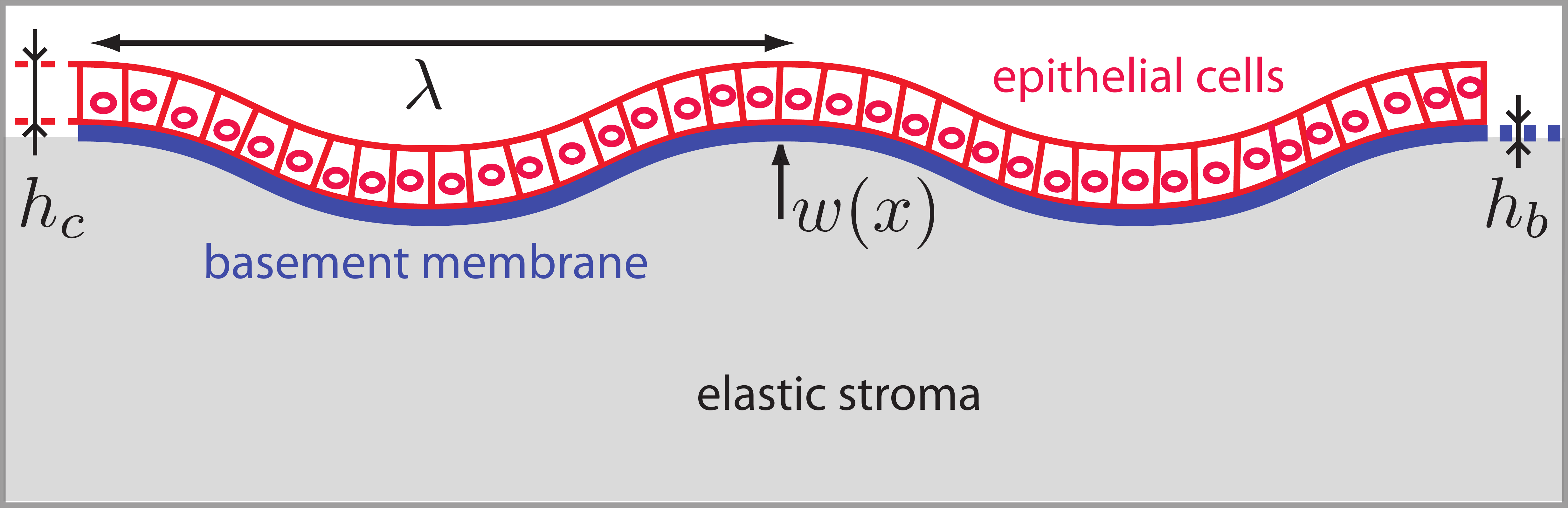}
\caption{Schematic description of an epithelial monolayer on top of a basement membrane and an elastic stroma. Crude estimates for the values of the  parameters are 
\cite{11} \cite{12}, $h_c = 10\mu m, h_b=1 \mu m, E_s 
= 400 Pa, E_c = 10^4 Pa, E_b=10^3 Pa$. }
\label{figure1}
\end{figure}

The intestine is a relatively simple organ: it is covered by a single layer of 
epithelial cells, which lay on top of the thin and relatively stiff basement membrane. 
The tube is surrounded by a soft connective tissue called the stroma.
We adopt here a three layer model for the intestine, sketched on Fig. \ref{figure1}. 

The underlying stroma is 
considered as an elastic medium of height $H$ and Young's modulus $E_s$. In practice, 
the thickness of this connective tissue can be considered as 
infinite as it is larger than the typical wavelength of the villi  
$\lambda \approx 100 \mu m$. The basement membrane is a thin elastic sheet of height 
$h_b \approx 1 \mu m$, with a bending modulus $K_b$ and an elastic modulus $E_b$.
The cell monolayer, has a thickness $h_c \approx 10 \mu m$, and an elastic modulus $E_c$. 
Considering the cell density as a constant, its bending 
modulus is $K_c = \frac{1}{9}E_c h^3_c$, which is much larger than the bending modulus 
of the basement membrane. The elasticity of the cell monolayer 
is dominated by curvature, and has no stretching energy, since the cells respond to strain by dividing 
or undergoing apoptosis over long time scales. We also consider that the cell monolayer 
can glide on the basement membrane.

We first give a simple argument based on energy considerations which 
gives good insight into the mechanism of the instability and allows us
to estimate orders of magnitude. 
The buckling instability is driven by the cell layer. When cells divide or die in a tissue, 
they exert stresses on their environment. The characteristic pressure is the 
homeostatic pressure $P_h$ \cite{9}, which is exerted on its surroundings by the tissue in its 
steady state, i.e., when cell division balances cell apoptosis.  Since the intestinal 
epithelium is a monolayer, the forces in the plane of the monolayer can be 
described in terms of a negative surface tension $-\gamma$ related 
to the homeostatic pressure by $\gamma =  P_h h_c$.
At the lowest order, the elastic energy of an undulation $ w(x) = w_0 cos(q x) $ of the cell 
monolayer reads :

\begin{equation}
\small
{\cal{E}} = \frac{K_c}{2} w_0^2 q^4 - \frac{\gamma}{2} w_0^2 q^2 +
\frac{E_s}{3} w_0^2 q
\end{equation}
\normalsize

Buckling occurs when the energy of a finite wave vector undulation
becomes negative i.e. if the tension $\gamma$ is larger than the critical value 
$\gamma_c = (3 K_c E_s^2)^{\frac{1}{3}} $.
The wavelength of the unstable undulation at threshold is  then $\lambda_c = 
2\pi \left(\frac{K_c}{E_s}\right)^{\frac{1}{3}} \approx 90 \mu m$, close to the observed \emph{in vivo} wavelength (cf. \ref{figure1}). Remarkably, analytical solutions obtained by minimizing the full elastic energy 
above threshold, for an arbitrary pressure, retain the same wavelength. 
The critical homeostatic pressure for villi formation is typically 
$P_{hc} = 1400$ Pa.m. This value is compatible with a homeostatic 
pressure in the range $10^3-10^4$ Pa \cite{26}.  One can therefore expect
the intestinal lining to buckle under the pressure of the epithelium.

Surprisingly, animals of very different sizes (mice and humans for instance), have villi 
of equivalent dimensions. This makes sense in our description, the size and wavelength
are only dictated by elastic considerations, and the elastic  parameters are 
roughly constant in most mammals. 
 
The previous energetic approach reproduces accurately the undulations of the small
intestine. The architecture of the colon is visually very different \cite{15}: there are 
no villi but only crypts extending into the stroma. 

This asymmetry prompted us to take into account the non-uniform division rates along the villi. 
\emph{In vivo}, cells multiply in crypts from stem cells and undergo apoptosis 
at the tip of the villi. They constantly flow from 
the crypts to the villi, differentiating along the way. The time scale for complete
renewal of an epithelium is 3 days in the intestine, and 7 days in the colon.
We do not wish to enter here in the molecular details, 
we will suppose that the non-uniform cell division can by modeled by a growth 
rate depending on the curvature of the cell monolayer, a hypothesis which has biological grounds \cite{30}.

We use as a reference state the homeostatic state $\gamma = \gamma_0$
where the monolayer is flat and reaches a steady state so that the cell division rate $k_d$ balances the cell apoptosis rate $k_a$, then we expand our equations around this state.  Since the monolayer is polar, the apical and basal sides are not equivalent and a linear coupling with curvature is allowed :
$ k_d - k_a = - \xi (\gamma - \gamma_0) + \alpha \nabla^2 w $. 
The coefficients $\xi$ and $\alpha$
measure the dependence of the growth rate on pressure and 
on curvature respectively. They are positive so that the 
growth rate decreases with pressure and is larger in the crypts.
Assuming that the cell density in the monolayer is essentially constant, 
the cell conservation equation reads: 

\begin{equation}
\small
\label{conservation}
\nabla . \textbf{v} = k_d - k_a = - \xi (\gamma - \gamma_0) + \alpha \nabla^2 w  
\end{equation}
\normalsize
The pressure gradient in the monolayer is balanced by the friction force with the
basement membrane: $\zeta \vec{\textbf{v}} = - \vec{\nabla} \gamma$, where $\zeta$ is a
friction constant. These two equations give the pressure in the cell monolayer 
as a function of the undulation amplitude. For small friction coefficients 
$\gamma = \gamma_0 - \alpha \zeta w$ while for large friction coefficients 
$\gamma = \gamma_0 + \frac{\alpha}{\xi} \nabla^2 w$.
It is difficult to obtain precise values of the parameters $\zeta$, $\xi$ and $\alpha$ 
and therefore to decide which regime is more relevant. We have 
performed calculations in  both regimes, and observed  the same physical 
behavior. In the following, we present calculations 
in the small friction regime, and define $p = \alpha \zeta$. 
 
\begin{figure}[!h]
\centering
	\includegraphics[width=7cm]{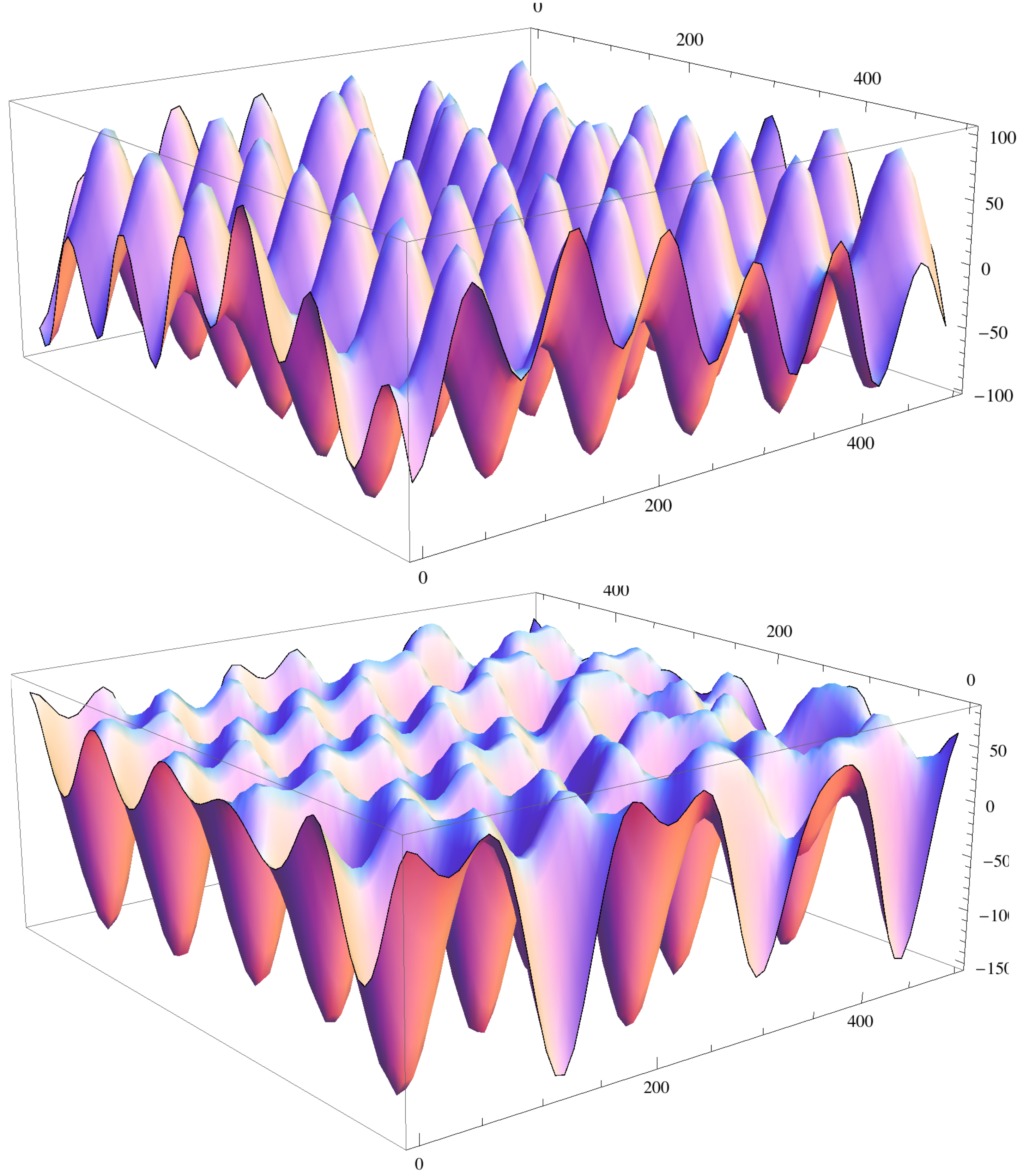}

	\caption{(a) Small intestine morphology, showing developed villi, and (b) colon morphology, showing only crypts. Units are micrometers.}
	\label{figure2}
\end{figure}

For large values of  $\alpha$, the crypts and villi become asymmetric. Villi 
are flattened, since the pressure is locally lowered by apoptosis, and crypts are deepened, 
since the pressure is locally increased by division (Fig. \ref{figure2} ).

 We start the analysis with $\alpha = 0$, in order to study the morphology of the small intestine. 
A large $\alpha$ is then included to explain specificities of the colon. 
Our main result is a phase-diagram which presents the possible \emph{in vivo} 
morphologies of the intestinal tube.
 
For each point in the monolayer, we define a displacement vector  ${\bf u}=(u_1,u_2,w)$,
where $w$ is the vertical displacement of the interface and $(u_1,u_2)$ 
the lateral displacement. We obtain the steady state shape and structure of the villi
from the force balance on the basement membrane. We introduce in the force balance a local drag force proportional to the local velocity of the 
membrane, which controls the relaxation of the membrane toward its equilibrium shape. 
The actual dissipative force is certainly non local and more complex but as we are 
interested only in the steady state shape, this provides a convenient way to relax, in 
our numerical solutions, the system toward its steady state configuration. 
The basement membrane is described by the classical Foppl-Von-Karman \cite{17} 
equation, which applies to thin rigid plates at moderate deflections. We consider for simplicity a membrane with vanishing Poisson modulus.

After rescaling all lengths by the thickness of the basement membrane $h_b$ and time, the 
equations of motion of the basement membrane are :
\small
 \begin{eqnarray}
 \label{force}
&&\frac{\partial w}{\partial t}  = - \Delta^2 w - \gamma \nabla^2 w - f_{el,z} \nonumber\\
&+& 9 \left(\frac{h_b}{h_c}\right)^{\frac{1}{3}} \frac{E_b}{E_c} \frac{\partial}{\partial x_\beta} \left(\frac{\partial w}{\partial x_\alpha}\left(\frac{\partial u_{\alpha}}{\partial x_{\beta}} + \frac{\partial u_{\beta} }{\partial x_{\alpha} } + \frac{\partial w}{\partial x_{\alpha}} \frac{\partial w}{\partial x_{\beta}}\right)\right) \nonumber\\
&&\frac{\partial u_\alpha}{\partial t} = 9 \left(\frac{h_b}{h_c}\right)^{\frac{1}{3}} \frac{E_b}{E_c}   \frac{\partial}{\partial x_\alpha} \left(\frac{\partial u_{\alpha}}{\partial x_{\beta}} + \frac{\partial u_{\beta} }{\partial x_{\alpha} } + \frac{\partial w}{\partial x_{\alpha}} \frac{\partial w}{\partial x_{\beta}}\right) \nonumber\\
&-& f_{el,\alpha}
\end{eqnarray} 
\normalsize
\normalsize

The various terms in equation (\ref{force}) describe the friction force, the 
curvature of the cell layer, the pressure exerted by the cells, the stretching force 
of the basement membrane, and the elastic force due to the stroma. The elastic 
force of the stroma ${\bf f}_{el}$ is proportional to the membrane displacement
${\bf u}$. Although we are not writing it for the sake of simplicity, we also took into account the nonlinear dependance of the curvature and tensions forces on $\vec{\nabla} w$, deriving them from their full Hamiltonian. This does not change the qualitative results, but raises the issue of which non-linear elasticity should be chosen for the stroma. This is beyond the scope of this paper and should be addressed in future work. The linear relation  for ${\bf f}_{el}$ is best written in Fourier space 
for a wave vector $\bf k$ as ${\bf f}_{el}=3 
\frac{E_s}{E_c} (\frac{h_b}{h_c})^{\frac{1}{3}} {\bf M}   {\bf u}$ where ${\bf M}$ is the  
$3 \times 3$ matrix given by
\small
\begin{equation}
 \left( {\begin{array}{ccc}
 (2k^2-k_y^2)/k & k_x k_y /k & 0  \\
 k_x k_y /k &  (2k^2-k_x^2)/k &0   \\
 0 & 0 & 2 k \\
 \end{array} } \right)
 \end{equation} 
 \normalsize
 
\begin{figure}[!h]
\includegraphics[width=8.5cm]{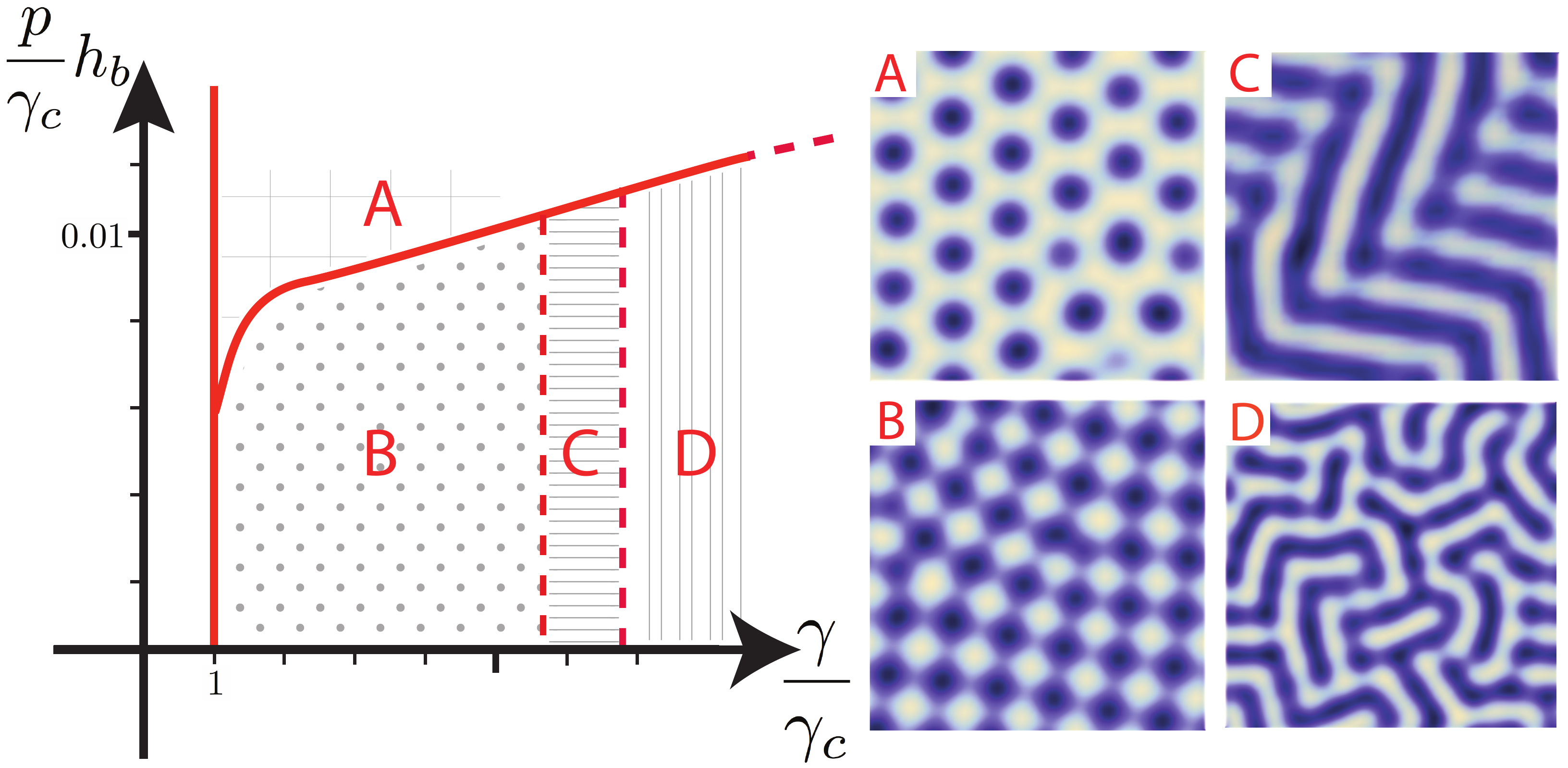}
\caption{\label{diagramme} Phase diagram showing the various possible villi organizations. Fig. A is colon-like whereas B, C and D are small-intestine-like, respectively finger-shaped, herringbones and labyrinth. The vertical axis is the coupling between cell division and curvature and the horizontal axis is the pressure exerted by the cell monolayer.}

\end{figure}

Whereas, the buckled state is a simple sine-like function in 1D, there are a variety of possible patterns in two dimensions \cite{19, 20, 21}. 
The three main morphologies are finger-shaped villi, herringbone, or disorganized labyrinth patterns. Following \cite{22}, we implemented a semi-implicit integration method. All products are done in real space, and all derivations in Fourier space, for maximal efficiency. As shown in Fig. \ref{diagramme}, we obtain a phase diagram regrouping different equilibrium solutions for characteristic physiological values of our parameters.  The parameter that we vary here is the buckling pressure exerted by the monolayer, on the horizontal axis. 

At low buckling pressure, finger shapes are stable, but for large stresses, it becomes 
difficult to bend the membrane in all directions, and labyrinth or 
herringbone structures are favored. This has been observed experimentally for 
various inert thin film systems \cite{18}.
And indeed, villi are not always finger-shaped. They sometimes adopt complex 
folded patterns strikingly identical to the ones that we predict \cite{4} \cite{31}. 

To quantify the buckling pressure, a good proxy is the height of villi, as indicated by equation (\ref{eq8}). Although villi wavelength is roughly constant throughout the intestine, villi height is not. At the entrance of the intestine (near the stomach), villi are very long, but their size decreases gradually in more distal parts \cite{31}.  This leads us to conclude pressure decreases along the intestine, justifying the relevance of our phase diagram.

\begin{figure}[!h]
\centering
	\includegraphics[width=8.6cm]{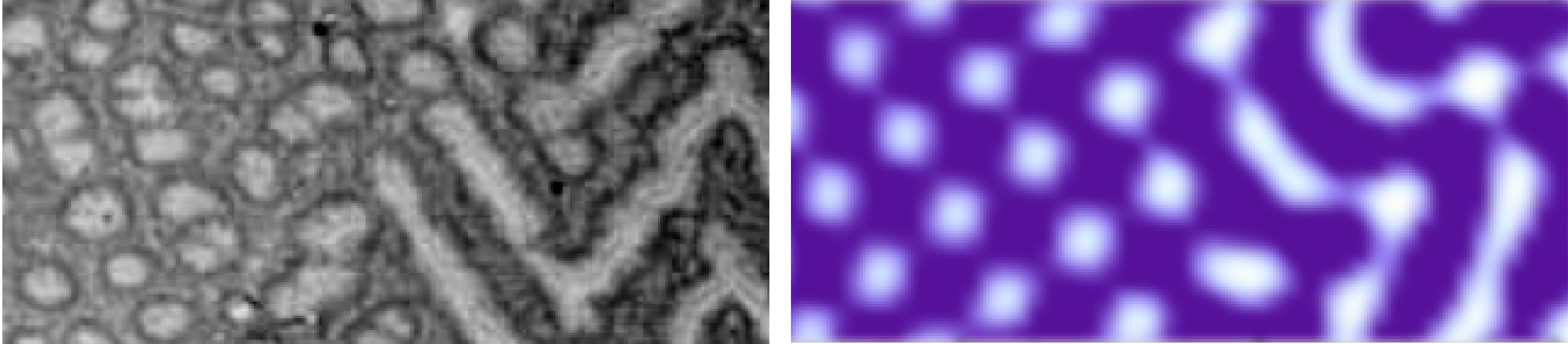}

	\caption{Transition between finger and herringbone villi, comparing a transverse cut (left) and our simulation (right). Pressure increases along the horizontal axis from left to right.}
	\label{figure4}
\end{figure}

As an example, Fig. \ref{figure4} shows a transition between finger-shaped and herringbone villi, \emph{in vivo} \cite{31}, and from our numerical integration when increasing the pressure along the horizontal axis.

We therefore expect that labyrinth-like villi are more likely to be found near the stomach, and that in distal parts, villi should be finger-shaped. Indeed, this is what physiological studies show \cite{4}. 

While we do not claim that this is the mechanism for the formation of villi
during embryogenesis, we argue that when the system reaches a steady state, 
stresses must be balanced, as described by our equations. In fact,  several experiments have shown that when increasing cell apoptosis for a short time \cite{24}, villi disappear, and in a matter of days, regain their initial equilibrium state.

Seeing villi as a product of a buckling instability could shed new light on intestinal pathologies. It should be noted that in our model, the elasticity of the basement membrane plays a crucial role, since it stabilizes the amplitude of the folds. In coeliac diseases, atrophy of the villi correlates with an increase of the apoptotic index, but also with an increase of the basement membrane thickness \cite{28}. This translates into a lower pressure, and a higher energetic cost to bend the membrane, which can cause the pressure to decrease below the buckling threshold, and thus the abrupt disappearance of villi.
Conversely, if the basement membrane breaks, which is a necessary event in colon cancer, this model predicts uncontrolled outgrowth of the intestinal lining, similar to polyp formation

In addition to this equilibrium state, there are secondary buckling events in crypts that result in a 
much higher crypt to villi size ratio than described here. This mechanism has been investigated 
in Ref.\cite{14} from a physical point of view and we will ignore it in this work.

The colon can be studied by considering a high asymmetry in cell division rate due 
to curvature.  The pressure is then a function of the deformation $w$ as 
$\gamma \approx \gamma_0 - p  w(x,y,t)$. The epithelium renewal rate in the colon is about 
twice slower than in the small intestine, suggesting that the stresses exerted are also lower. 
Also relevant is the fact that CDX1 and CDX2 levels are higher in the colon \cite{15}, 
and that these protein inhibit $\beta$ - catenin, thus favoring earlier differentiation.
Physically, this would mean that high stress regions are more localized in colonic crypts than intestinal crypts, and that the asymmetry parameter $p$ is higher in the colon. 

Integration of these new equations for several values of $p$ demonstrate, as shown Fig. \ref{figure2} and Fig. \ref{diagramme} A and B, a transition between intestinal and colonic morphologies. This transition is sharp, as observed \emph{in vivo} at the intestinal-colonic junction. This is consistent with the following analytical calculation made for small values of $p$ in a one-dimensional geometry.

In one dimension the monolayer deformation amplitude follows the equation 
\begin{equation}
\small
0 =  K_c \Delta^2 z + \gamma_0 \Delta z - E_b h_b \vec{\nabla}.((\vec{\nabla} z)^2 \vec{\nabla z}) + f_{el} - p z (\Delta z )
\end{equation}
\normalsize
For small values of $p$, we expand the solution as a sum of two cosine functions: 
$h(x) = h_0 + h_1cos(qx)  + h_2 cos(2qx)$, with $q = \frac{E_s}{\gamma_c}$ and  
identify  the  terms in $cos(qx)$ and $cos(2qx)$. The amplitudes are such that  
$h_2^2 \ll h_1^2$ and $p h_2 h_1\ll h_1^3$. Within this approximation, the value of $h_1$ 
is unchanged compared to the  symmetric $p=0$ case :
\begin{equation} 
\small
\label{eq8}
h_1^2 = \frac{2 \gamma_c^3}{E_s^2 E_b h_b}(\frac{\gamma_0}{\gamma_c}-1) 
\end{equation}
\normalsize
The non-symmetric term $h_2$ is negative, in agreement with what is expected: 
when subtracting $cos(2x)$ to a larger $cos(x)$ term, the troughs are deepened and 
the crests flattened. 
\begin{equation}
\small
h_2 =  -  \frac{ 3 p h_1^2}{8 ( 3 \gamma_0 - \gamma_c)} < 0
\end{equation}
\normalsize

Our numerical solutions predict that intestinal villi are 
organized in a square lattice (as can be proven analytically by energy considerations,
following \cite{21}), whereas colonic crypts are organized on an hexagonal lattice as  
seen on Fig. \ref{diagramme} (in agreement with weak crystallization theories \cite{33}). Although the system is noisy, this is strongly hinted \emph{in vivo} by the fact that transverse cuts of the small intestine do show square-shaped villi, whereas crypts are round and have a higher number of neighbors.

This simple buckling model predicts a full phase diagram explaining most of the 
structures observed in the small and large intestines. This suggests a deep connection 
between tissue architecture and the stresses produced by the dividing cells. Intestinal 
shape and renewal are at least partially controlled by a mechanical balance,
which is disrupted in the case of intestinal diseases. Since it has been proven that an
excess of mechanical pressure could, in itself, induce colonic cancer \cite{27}, a detailed understanding of this balance is of great interest for cancer research.

We thank S. Fre and M. Huygue for discussion and for showing us in vivo samples, and A. Berg\`es for help with the manuscript.

\end{document}